# Sociotechnical Considerations for Accessible Visualization Design

Alan Lundgard* [†]  Crystal Lee* [‡]  Arvind Satyanarayan[§]
MIT CSAIL          MIT HASTS         MIT CSAIL

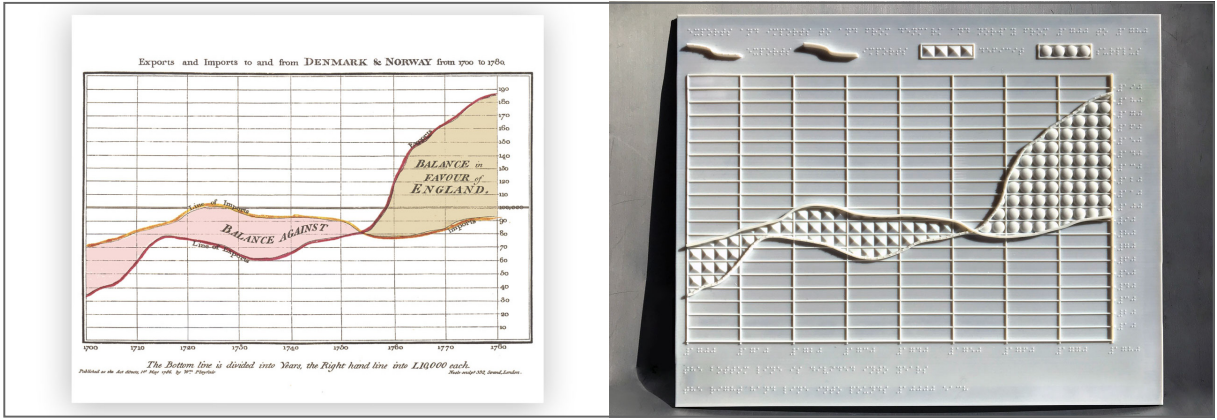

Figure 1: As part of an inclusive design workshop at the Perkins School for the Blind, we created a 3D printed tactile translation of a time-series chart by William Playfair. In this paper, we show how these one-to-one translations, while based on existing best-practice guidelines for tactile graphics, can be pedagogically ineffective and incur prohibitive costs.

**ABSTRACT**

Accessibility—the process of designing for people with disabilities (PWD)—is an important but under-explored challenge in the visualization research community. Without careful attention, and if PWD are not included as equal participants throughout the process, there is a danger of perpetuating a vision-first approach to accessible design that marginalizes the lived experience of disability (e.g., by creating overly simplistic "sensory translations" that map visual to non-visual modalities in a one-to-one fashion). In this paper, we present a set of sociotechnical considerations for research in accessible visualization design, drawing on literature in disability studies, tactile information systems, and participatory methods. We identify that using state-of-the-art technologies may introduce more barriers to access than they remove, and that expectations of research novelty may not produce outcomes well-aligned with the needs of disability communities. Instead, to promote a more inclusive design process, we emphasize the importance of clearly communicating goals, following existing accessibility guidelines, and treating PWD as equal participants who are compensated for their specialized skills. To illustrate how these considerations can be applied in practice, we discuss a case study of an inclusive design workshop held in collaboration with the Perkins School for the Blind.

**Index Terms:** Human-centered computing—Visualization—Visualization design and evaluation methods; Human-centered computing—Accessibility—Accessibility technologies

*authors contributed equally
[†]e-mail: lundgard@mit.edu
[‡]e-mail: crystall@mit.edu
[§]e-mail: arvindsatya@mit.edu

## 1 INTRODUCTION

History provides many instructive examples of faulty information systems designed for people with disabilities (PWD).[1] For example, early approaches to designing tactile reading systems for blind people frequently used embossed letters that attempted to translate the alphabet's visual contours directly into tactile shapes. In the 18$^{th}$ century, schools favored these systems because sighted teachers did not require special training to read the embossed letters, and because it allowed blind and sighted readers to read the same texts [28].

However, drawing on his own experience with blindness, Louis Braille—the inventor of the ubiquitous reading system—knew that raised dots were easier to read than embossed letters because they provided more tactile distinction. The eventual success of Braille's system over embossed letters is a parable for solutions that are designed by and for PWD. Since PWD are better informed about their lived experiences than designers who are not disabled, they are also in the best position to design and engineer sustainable solutions. Embossed letters were ineffective because they directly translated visual into tactile systems *without* centering a blind reader's experience.

This story resonates today as technologists extol the potential for web interfaces and 3D printing to improve the accessibility of data visualizations. Since the 1990s, visualization and Human-Computer Interaction (HCI) researchers have developed tactile interfaces that claim to make information more accessible to blind people by allowing users to touch, rather than see, the data [12, 22, 24]. However, few of these projects extensively consult blind people, which means that they often become what disability design expert Liz Jackson calls *disability dongles*: a "well intended, elegant, yet useless solution to a problem we [PWD] never knew we had" [23]. Developments in HCI and tactile graphics have previously converged in Assistive Technology (AT) research [7, 26, 32], but visualization researchers have not yet substantively engaged with the developments in tactile information systems, well-established accessibility guidelines, or with the perspectives of PWD.

---

[1]In this paper, we use "people with disabilities" as suggested by the ACM SIGACCESS guidelines [18]. In the United States, scholars have primarily used *person-first language* (e.g., "people with disabilties"), as opposed to *identity-first language* (e.g., "disabled people"). Our paper generally uses this convention, although we also recognize that some people prefer identity-first language (e.g., "blind people").

To begin addressing these issues, we present a set of *sociotechnical* considerations for accessible visualization design. In our analysis, we emphasize that there are both *social* and *technological* constraints on the accessible visualization design space. We demonstrate how our considerations—informed by work written by and for PWD in disability studies, AT, and design history—can be applied to future accessibility research. We also critique current research that emphasizes a vision-first approach to making tactile graphics.

## 2 BACKGROUND & RELATED WORK

Our work brings together disability studies, research on tactile information systems, and participatory design methods to articulate a set of sociotechnical considerations for accessible visualization design.

### 2.1 Disability Studies

In the United States, disability studies is an interdisciplinary field that seeks to untangle the different political, intellectual, and cultural dimensions of disability in society. In this literature, scholars distinguish between two models of thinking about disability [40]. In a **medical model,** the disability diagnosis is tied directly to an individual's physical or psychological state, and the prognosis focuses on curing or managing the disability until it disappears as much as possible (e.g., a deaf person who uses cochlear implants) [17]. By contrast, a **social model** of disability distinguishes between an *impairment* (a physical or psychological abnormality) and a *disability,* which describes a form of social, economic, and political exclusion perpetuated against people who have some impairment [35]. In the social model, people are disabled by structural and environmental factors, not by their bodies [41].

While these two models are not mutually exclusive, and can be considered alongside personal, environmental, and other contextual factors [2], it is important for visualization researchers to understand the social and medical models because each can be leveraged towards different kinds of accessibility research. The medical model, for example, clearly articulates an actionable problem that can be measured and fixed, which can be practically useful to AT researchers. However, AT research that focuses exclusively on the medical model often stigmatizes PWD by attempting to "fix" all impairments and differences through medical intervention, even when the condition does not cause pain, illness, or death (e.g., deafness) [14]. In particular, the medical model misses how an impairment *develops* into a disability. For example, in a society where everyone communicates through sign language, deafness may not be perceived as a disability at all, but may instead be a source of cultural pride. In a hearing society, however, deafness may be seen as a disability that requires accommodation through interpreters or medical intervention [27, 30].

The social model of disability, by contrast, shifts the focus from "cure to care," in which the goal is not to fix the impairment, but rather the ways that society treats PWD [44]. For example, rather than building a better wheelchair that allows its user to climb stairs, proponents of the social model look instead to dismantling the legislative or bureaucratic barriers to installing wheelchair ramps in public places. While the social model cannot explain abnormal biological functions, it can be crucial for driving the adoption of assistive technologies and ensuring that the technologies address the infrastructural barriers that PWD experience in their everyday lives.

### 2.2 Tactile Information Systems

Tactile systems like braille displays, 3D models, and embossed maps have been a mainstay of blind education [42]. As such, there is a vast literature by computer scientists, education professionals, and organizations for the blind that explore how these tools can be put into practice [4, 16, 39]. Education researcher Lucia Hasty, for example, has developed principles of graphical literacy that show how blind and sighted students learn differently [19]. Sighted learners absorb information **whole to part**, where they see the whole picture simultaneously, and understand the different visual encodings in relation to each other (e.g., the size and color of one bar compared to another). Tactile learners, however, must approach tactile graphics **part to whole** by touching individual parts of the graphic processually. They then put together each piece of information in a sequence to view the graphic as a whole. Given the differences between these approaches, successful tactile representations should ensure that it is easy to access the information sequentially, and that each tactile element is easily and quickly distinguishable. Braille is a prime example of a successful tactile representation. Each braille cell is easy to scan quickly and sequentially (compared to embossed lettering, where a user must try to synthesize each letter individually—a relatively slow process—before moving onto the next letter). Furthermore, printed braille has many well-established contractions, which make it even faster to scan through texts. Drawing on braille's design successes and guidelines, an important area for future work is developing an analogous system for visualizations (i.e., one that permits fast and effective information access, and does not simply translate the visualization into a tactile representation in a one-to-one fashion). Combining these insights with guidelines on tactile graphics from organizations like American Printing House for the Blind (APH), Braille Authority of North America (BANA), and World Wide Web Consortium (W3C)'s Web Accessibility Initiative (WAI), provide a powerful basis on which to begin this research [1, 3, 5]. Building on these existing AT guidelines is an excellent example of how the social model of disability can inform practice [27].

### 2.3 Participatory Design Methods

At each stage of conducting a study, researchers should be attentive to the power dynamics between themselves and their study participants. Historically, ignoring these dynamics has exposed already marginalized communities to long-term psychological and physiological harm (e.g., the Tuskegee syphilis experiment) [10]. In particular, developments in AT have often relied upon using PWD as test subjects for technologies that were later transformed into more profitable ventures intended for able-bodied people [29]. To mitigate situations like these, researchers in many disciplines have developed ways of working *with* (rather than *on*) marginalized communities [11, 13, 31]. Generally, these community-based or participatory research methods emphasize collective inquiry in which study participants are considered co-researchers. Far from simply being a *subject* of research, these participants help define the design problem and contribute to methods, data collection, analysis, and publication [27, 32, 33]. These methods go beyond user-centered design [34] to articulate more clearly what the stakes are in a research project: who participates with whom in what? Who are the intended beneficiaries of a project, and how do they accrue these benefits? By re-centering PWD in the design process, participatory methods can be useful for historically marginalized communities because they break down the separation between those who are doing the design and those who are being designed *for*.

This work also attempts to avoid the problems associated with *parachute research*, a phenomenon in which researchers—particularly those from wealthy universities—drop into a community, make use of local infrastructure and expertise, and then disengage from the community altogether after publishing results in a prestigious academic journal [20]. This kind of research is harmful to the communities who take the time and resources to help facilitate academic research without reciprocal benefits. This can have a disproportionately negative impact on disability communities, who already face many barriers to participating in public life. To address these problems, scholars in AT, disability studies, and design have emphasized how researchers need to consider their participants as active agents with ideas and goals that may conflict with those of the researcher. This should not be seen as an obstacle to research; rather, it provides new opportunities for collaborative design [21, 38].

## 3 SOCIOTECHNICAL CONSIDERATIONS

In this section, we introduce a set of sociotechnical considerations for accessible visualization design. These considerations are informed by related work, our own decade-long engagement with the blind community, and a case study of an inclusive design workshop. Each consideration begins by explaining its social and technological aspects, and concludes with questions that researchers and designers should consider when collaborating with disability communities.

**Non-Intervention.** At each step of the design process, researchers should consider whether any technological intervention is appropriate at all [6]. This is especially true of designing for and with PWD because well-meaning interventions may worsen the very situations they are intended to help [23]. Design processes are actively harmful when they exhaust collaborators' time and resources without adequate compensation and reciprocity, or when they are simply band-aid technological solutions to infrastructural problems. Is there an equally viable low-tech or no-tech solution? Might the technological intervention result in more harm than the problem it is meant to address? Does the technology solve a computationally tractable version of the problem, or does it address an actual need?

**Research & Design.** If technological intervention is appropriate, researchers should carefully evaluate the goals of their project. In HCI, there is a well-known tension between doing "research" and doing "design" [36, 43]. On the one hand, research has the goal of creating new knowledge, but this may not be the best solution for a user's immediate needs. On the other hand, design should satisfy those needs, but the solution may lack research novelty. The two are not mutually exclusive, but they do prescribe different methods and goals (e.g., a research publication versus an adoptable design solution). This tension is especially apparent when designing for and with PWD. Preliminary interviews may reveal that the best design solutions to a user's needs will not count as publication-worthy material. Is the proposed solution addressing those needs, or is it a novel research contribution? Is there a solution that addresses both?

**Participatory Methods.** Taking a participatory approach to research and design is a key tenet of accessibility. This involves including stakeholders in the design process from the outset, during need-finding, problem definition, prototyping, publication, and dissemination [27]. Participatory design aims to include users as equal participants in the design process, as opposed to merely verifying the usefulness of a solution via user studies [15]. Inclusive design, as complementary to participatory design, places emphasis on empowering as many people as possible, often with a focus on removing barriers for people who have physical or cognitive disabilities [32]. Inclusive approaches may involve consideration of the *interdependence* between researchers and stakeholders (i.e., collapsing the distinction between who is doing the research, and *for whom* that research is being done) [7, 8]. Who are the stakeholders in a research project? Are they all recurring and equal participants throughout the design process? Are the research and design tools themselves accessible [25]?

**Communicating Expectations.** Because HCI and AT research projects can often be one-off or proof-of-concept prototypes that are no longer maintained after publication, it is especially important for researchers to communicate their intentions, expectations, and capabilities clearly to all collaborators. This may include the intended duration of the project, the quantity of available resources, and the project goal—whether it be a research publication or adoptable design solution. This is especially important for technical projects that require maintenance or engineering support beyond the duration of the project. How long will a project be maintained, and what are the expected contributions of each collaborator? If the goal is academic publication, has there been a discussion about author credit and order? If the goal is a marketable prototype, has there been a discussion about equitable compensation and intellectual property?

**Time & Compensation.** It is generally good practice to be sensitive and respectful of all collaborators' time, but this is especially true when designing for and with PWD. Like everyone, PWD are busy, but they must also contend with additional barriers to mobility, access, and employment. This makes the time that they spend on a design project especially valuable. As with any specialist, PWD should always be compensated at a rate commensurate with their specialized skills, such as the ability to read braille or use a screen reader. Additionally, it may be difficult to find and recruit PWD for collaboration, and this should be reflected in their rate of compensation, meaning a rate that is greater than that of an average user study participant (e.g., in some cases $35.00 per hour) [9, 37]. How much time does each participant have to contribute to the collaboration? Would it be better to not participate if it is not possible to contribute meaningfully or to bring the project to fruition? Are participants with disabilities being compensated adequately and fairly?

**Accessibility Guidelines.** Researchers should familiarize themselves with the accessibility guidelines that have been developed by major organizations like the APH, BANA, and the W3C's WAI [1, 3, 5, 16]. While these guidelines are not complete and are sometimes contested, researchers should use them to steer and expand their inquiry. For example, accessibility guidelines for 3D printed tactile graphics might build upon the BANA guidelines, and accessible web-based visualizations might integrate Accessible Rich Internet Application (ARIA) attributes developed by the W3C. Are there existing standards and best practices relevant to the design problem? If so, does the proposed design solution adhere to these standards? If not, in what way might the proposed design solution integrate with, or build upon, this work?

**Technology Access.** Designing for and with PWD almost always involves securing access to specialized materials and technologies, from low-tech solutions (e.g., tactile tape, puff paint, wax strips) to high-tech devices (e.g., screen readers, refreshable braille displays, embossing and 3D printers). Access to these technologies will circumscribe the space of feasible design solutions, and so particular approaches should be chosen carefully to fit within these constraints. A high-tech approach may permit finer-grained, more durable products, and it may yield more sensational results, but securing access to these technologies can be prohibitively costly. A low-tech approach may be more readily available to the user on a daily basis, but may convey information more coarsely or unreliably. Even ostensibly democratizing technologies, such as 3D printing, may be hard to come by, and the interfaces to operate these technologies may not be accessible themselves. Does the problem require a high-tech approach, or will a low-tech approach work just as well, if not better? Does the design require one-time access to an expensive piece of technology, or repeated, frequent access by the user?

**Technology Resolution.** In addition to constraints imposed by access to specialized technologies, researchers also need to ensure that those technologies can encode information effectively. By drawing an analogy with high-fidelity audio and display resolution, we use "resolution" to describe how well a particular medium can encode and convey detailed information. This is especially important for web visualizations, which are typically not screen reader compatible, and for braille, which requires standardized height and spacing, and cannot be resized to fit a particular area. For example, with 3D printing, the size of the printer bed may constrain the amount of braille that can be printed in one line, and the quality of the printing material may change how long a braille reader can interact with the object. Plastic filaments used on consumer 3D printers, for example, can create abrasive surfaces that make it uncomfortable for braille readers to use for a long period of time. A medium's resolution depends on various hardware and software limitations, and on material and social constraints. What degree of resolution is appropriate to the user's needs, and which technologies can be used to achieve it?

## 4 CASE STUDY: THE PERKINS SCHOOL FOR THE BLIND

To illustrate the benefits of approaching accessible visualization design from a disability studies and AT perspective, and to ground our sociotechnical considerations in a concrete example, we present a case study from our own work. We emphasize that this is not an ideal case, but a *problem case* exemplifying some of the pitfalls articulated in our set of considerations. Throughout this section, we refer back to these considerations in parentheses.

The authors took part in an inter-semester inclusive design workshop that featured a collaboration between MIT and the Perkins School for the Blind. The express agreement of this collaboration was to make design interventions that would address the needs of the Perkins School students and staff **(Non-Intervention)**. While the Perkins School had previously collaborated with other institutions and entrepreneurs, our collaboration involved multiple need-finding visits to the school's campus during which we toured the facilities and participated in discussions with occupational therapists, teachers, and technologists. This afforded opportunities for design iteration and feedback **(Participatory Methods)**. Of prior non-participatory collaborations, the President of the Perkins School noted that:

> *"These entrepreneurs come to us with their finished prototypes, but they haven't talked to very many blind people in the process before they've put in all the effort to create the prototype."*

In such cases, the President suggested that many of the resulting prototypes were relatively useless to their community, reflecting a sighted designer's idea of what a blind person *could* want, and not what they *actually* needed. By taking a participatory approach, the workshop was meant to evade this pitfall. However, we identified an ongoing tension between doing "research" and doing "design" which made clear that participatory methods are necessary but insufficient for framing design interventions. Although we did not enter the workshop intending to produce academic research, the workshop organizers placed an emphasis on developing a novel research product. They encouraged us to "think beyond" satisfying the commonplace and bespoke needs of the Perkins School students and staff towards new, more generalizable solutions **(Research & Design)**. To relieve this tension, we opted to work on a project that engaged with the immediate needs of blind students, but also had potential for research novelty—3D printed tactile graphics. One assistive technologist, for example, noted that prior approaches to tactile graphics in educational settings were ineffective:

> *"I have not found a tactile graphic solution that works. There's usually some sort of compromise. You might get the general outline of something, but you don't necessarily get the information that is conveyed... Now, if there was a new student, I would turn to 3D printing to create 3D visualizations."*

Because tactile graphics are often created using embossing printers that are limited in the number of printable dots-per-inch, direct translation of a high resolution visual image to a lower resolution tactile print leads to information loss. This can create tactile ambiguities that were not present in the original visual image. For example, in a paper embossed tactile graphic, the intersections between two lines in a multi-line plot would not be distinguishable through touch, as in the original Playfair chart (Figure 1). 3D printing permits higher resolution encodings because there is an extra printable dimension, which helps eliminate the ambiguities common to paper embossing **(Technology Resolution)**. Accordingly, 3D printing is often touted as a democratizing technology in terms of affordability and access, but the more affordable models (under $800) also have the lowest resolution and smallest printable dimensions. These constraints on print resolution may be incompatible with constraints imposed by existing accessibility guidelines. For example, BANA gives guidelines for the spacing between braille dots because braille becomes illegible when it is spaced too widely or narrowly [1]. Due to their low resolution and small printable dimensions, prototypes of the Playfair translation printed using the affordable 3D printers failed to meet the BANA guidelines **(Accessibility Guidelines)**. Thus, access to an extremely expensive commercial 3D printer (roughly $330,000) became a prerequisite for translating the detailed visualization into a high-resolution tactile graphic **(Technology Access)**. Compared with paper embossed graphics, the Perkins School students expressed a strong preference for the 3D printed graphics because of their higher resolution. This is a worthwhile research insight, but commercial 3D printing was not a design solution that was adoptable beyond the duration of the workshop and should have been communicated as such **(Communicating Expectations)**.

The collaboration between MIT and the Perkins School was well-meaning and it was intended to be mutually beneficial. As such, the Perkins School students and staff were not compensated for their valuable time and expertise, even though they spent part of their workdays to meet with us **(Time & Compensation)**. While this arrangement may have been equitable had we collaboratively produced solutions that actually satisfied the needs of the students, it was not clear that the workshop benefited the Perkins School students and staff as much as it benefited the participants from MIT. Media scholar Mara Mills has documented the many ways that disability has been used as a pretense to develop innovations that are primarily for publicity, often without giving back to PWD in a substantial way [29]. While this was in no way the workshop's intent, it may have been its predominant outcome: an opportunity for able-bodied researchers and designers to engage with PWD and to generate publicity for both institutions (both of which had reporters on-hand). For us, however, the workshop also afforded a valuable lesson for guiding future research and design that avoids the pitfalls of parachute research. Put succinctly, successful participatory design cannot be achieved within an accelerated time frame. Collaborations with PWD should support longer-term engagement through equitable compensation for each participant's time, and the goals of the design process should be well-scoped to account for each participant's availability and access to relevant technologies.

## 5 CONCLUSION & FUTURE WORK

In this paper, we contribute a set of sociotechnical considerations for accessible visualization design. Visualization research has largely focused on addressing the technological and perceptual barriers to the effective visual encoding of information. However, due to the many ways in which barriers to access are manifested, we emphasize that there are both *social* and *technological* constraints on the accessible visualization design space. While new technologies such as 3D printing and web interfaces afford many opportunities for future work (e.g., developing more effective tactile representations, guidelines for 3D printing, and screen reader compatible visualizations), existing work on braille and tactile graphics provide a guide for conducting this research successfully and inclusively. Drawing on these examples, our considerations are intended to help guide researchers away from *parachute research*, overly simplistic vision-first approaches (e.g., *disability dongles*), and towards design practices that avoid the pitfalls of well-meaning but insufficient collaborations with PWD (e.g., embossed lettering). When pursued with careful attention, there will be many exciting opportunities for collaboration between disability, AT, and visualization communities.

### ACKNOWLEDGMENTS

We thank the students and staff from the Perkins School for the Blind, the workshop organizers, and the MIT Assistive Technology Information Center. We also thank Michael Correll, Richard Ladner, and Meredith Ringel Morris for their feedback on earlier drafts.